\documentclass[journal]{IEEEtran}
\makeatletter
\def\ps@headings{%
\def\@oddhead{\mbox{}\scriptsize\rightmark \hfil \thepage}%
\def\@evenhead{\scriptsize\thepage \hfil \leftmark\mbox{}}%
\def\@oddfoot{}%
\def\@evenfoot{}}
\makeatother
\usepackage{xcolor}
\usepackage{xpatch}
\usepackage{subfigure}
\usepackage{colortbl}

\usepackage{graphicx}  
\usepackage{url}       
\usepackage{bm}

\usepackage{amsmath}   
\usepackage{cite}
\usepackage{amsfonts,amssymb}

\usepackage{stfloats}

\usepackage{cases}
\usepackage{algorithm}
\usepackage{multirow}
\usepackage{algorithmic}
\usepackage{stfloats}
\usepackage{epstopdf}
\makeatletter

\newcommand{\Rmnum}[1]{\expandafter\@slowromancap\romannumeral #1@}
\makeatother

\newcommand{\ls}[1]
    {\dimen0=\fontdimen6\the\font
     \lineskip=#1\dimen0
     \advance\lineskip.5\fontdimen5\the\font
     \advance\lineskip-\dimen0
     \lineskiplimit=.9\lineskip
     \baselineskip=\lineskip
     \advance\baselineskip\dimen0
     \normallineskip\lineskip
     \normallineskiplimit\lineskiplimit
     \normalbaselineskip\baselineskip
     \ignorespaces
    }

\begin{document}
\title{Reconfigurable-Intelligent-Surface Assisted Orbital-Angular-Momentum \\Secure Communications}
\vspace{10pt}
\author{\IEEEauthorblockN{Minmin Wang, \emph{Student Member, IEEE}, Liping Liang, \emph{Member, IEEE}, Wenchi Cheng, \emph{Senior Member, IEEE}, \\
Wei Zhang, \emph{Fellow, IEEE}, Ruirui Chen,  and Hailin Zhang, \emph{Member, IEEE}}\\[0.2cm]
\vspace{-10pt}


\vspace{-15pt}

\thanks{Minmin Wang, Liping Liang, Wenchi Cheng, and Hailin Zhang are with the State	Key Laboratory of Integrated Services Networks, Xidian University, Xi’an 710071, China (e-mail: minminwang@stu.xidian.edu.cn; liangliping@xidian.edu.cn; wccheng@xidian.edu.cn;	hlzhang@xidian.edu.cn).

Wei Zhang is with the School of Electrical Engineering and Telecommunications, The University of New South Wales, Sydney, NSW 2052, Australia
(e-mail: w.zhang@unsw.edu.au).
	
Ruirui Chen is with the School of Information and Control Engineering, China University of Mining and Technology, Xuzhou 221116, China (e-mail: rrchen@cumt.edu.cn).
}
}

\maketitle

\begin{abstract}
	As a kind of wavefront with helical phase, orbital angular momentum (OAM) shows the great potential to enhance the security results of wireless communications due to its unique orthogonality and central hollow electromagnetic wave structure. Therefore, in this paper we propose the reconfigurable-intelligent-surface (RIS) assisted OAM scheme, where RIS is deployed to weaken the information acquisition at eavesdroppers by adjusting the OAM beams pointed to the eavesdropper and artificial noise (AN) is applied to interfere with the eavesdropper, thus significantly increasing the secrecy rates of short-range secure communications. Aiming at obtaining the maximum secrecy rate, we develop the Riemannian manifold conjugate gradient (RMCG) based alternative optimization (AO) algorithm to assign much power to low-order OAM-modes and optimize the OAM beams direction with the programmable RIS, thus respectively enhancing and weakening the received signal strength at the legitimate receiver and the eavesdropper. Numerical results show that our proposed scheme outperforms the existing works in terms of the secrecy rate and the eavesdropper's bit error rate. 
\end{abstract}

\vspace{5pt}


\vspace{-0.2cm}
\section{Introduction}

\IEEEPARstart{D}{ue} to the broadcast and open properties of channels, wireless communication information is easily eavesdropped by illegitimate devices during transmission. This problem has promoted the research on  physical layer security (PLS) aiming at protecting communication information from being acquired by eavesdroppers. 

To achieve expected anti-eavesdropping results, the existing works on  PLS mainly focus on exploiting multiple-input multiple-output (MIMO) systems aided by joint transmit and receive beamforming, artificial noise (AN), and reconfigurable intelligent surface (RIS) \cite{Beamforming_Power_Design_for_IRS,IRS_assisted_untrusted_NOMA,AN_Aided_Secure_MIMO_via_IRS}. RIS has been studied for anti-eavesdropping in MIMO systems because it can reconfigure the propagation environment of wireless channels by flexibly adjusting the reflecting coefficients \cite{Beamforming_Power_Design_for_IRS}. 
The AN-assisted MIMO secure system verified the effectiveness of RIS in enhancing the security \cite{AN_Aided_Secure_MIMO_via_IRS}. 
However, due to the non-full rank channel matrices, it is very difficult for MIMO systems to dramatically enhance the secrecy results in line-of-sight (LoS) secure communication scenarios. Therefore, it is highly required to explore new technologies to achieve secure message transmission in LoS communications scenarios.

Fortunately, orbital angular momentum (OAM), which is associated with the helical phase fronts of electromagnetic waves, shows the potential to significantly increase the spectrum efficiency and enhance PLS of wireless communications \cite{Index-Modulation,OAM_Secure_High_Speed_Communication,Secure_Range_Dependent_Transmissio_With_OAM}. Signals can be transmitted in parallel among multiple OAM-modes due to the inherent orthogonality among different integer OAM-modes. Also, the hollow structure of OAM beams makes the difficulty for eavesdroppers to acquire confidential information. 
The secure and energy-efficient multidimensional coded modulation scheme for high-speed OAM communications was studied \cite{OAM_Secure_High_Speed_Communication}.  
The works of \cite{Secure_Range_Dependent_Transmissio_With_OAM} and \cite{Physical_Layer_Secure_Communication_Using_OAM} investigated secure information transmission with OAM beams generated by the designed transmitter. The range-angle dependent OAM secure transmission scheme with frequency-diverse array was proposed to decrease the possibility of OAM beams being eavesdropped \cite{ma2021OAMFDA}. The OAM-based physical layer secret key generation scheme was proposed  to improve the secret key capacity by exploiting the orthogonality among OAM modes \cite{zhang2022PLKOAM}. However, the existing studies on OAM secure communications primarily focus on designing antenna arrays and  secret key to improve the secrecy performance, where the OAM beams are not flexibly adjusted to weaken the signals obtained by the eavesdropper. Therefore, it remains an open issue of how to intelligently adjust OAM beams to enhance the security of short-range wireless communications.


To overcome the aforementioned issue, in this paper we propose a RIS-assisted OAM secure communication scheme, where the OAM beams direction is adjusted by RIS and AN is emitted by the legitimate transmitter to enhance the security of short-range wireless communications under eavesdropping. The secrecy rate maximization problem is formulated to jointly optimize the transmit power allocation and the RIS’s phase shift matrix. To solve the non-convex optimization problem with intricately coupled optimization variables, we develop the Riemannian manifold conjugate gradient (RMCG) based alternative optimization (AO) algorithm. Numerical results show that our proposed scheme can significantly enhance the secrecy results of short-range secure communications as compared with the existing works.

The rest of this paper is organized as follows. Section \uppercase\expandafter{\romannumeral2} gives the system model and formulates the secrecy rate optimization problem for the RIS-assisted OAM secure communication scheme. The RMCG-AO algorithm is developed to achieve the maximum secrecy rate in Section \uppercase\expandafter{\romannumeral3}. Section \uppercase\expandafter{\romannumeral4} presents the numerical results to evaluate the performance of our proposed scheme. Section \uppercase\expandafter{\romannumeral5} concludes this paper.

\begin{figure}[htbp]
	\centering
	\setlength{\abovecaptionskip}{0cm}
	\vspace{-10pt}
	\includegraphics[scale=0.45]{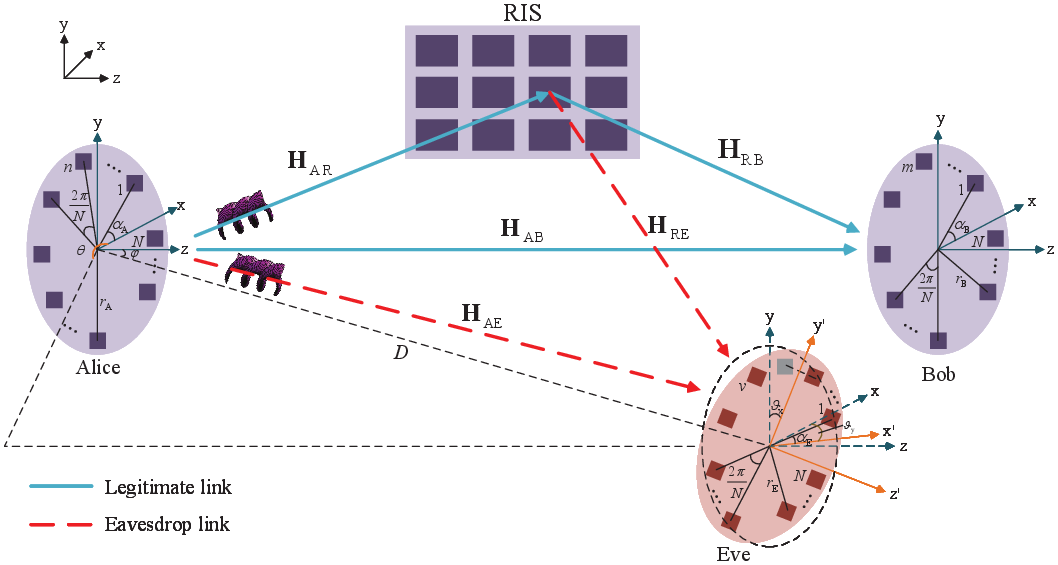}
	\caption{The RIS-assisted OAM secure communication system.} \label{fig: system model}
	\vspace{-12pt}
\end{figure}

\section{System Model and Problem Formulation}\label{sec:SYSTEM MODEL}
\subsection{System Model}
As shown in Fig. \ref{fig: system model}, we build a RIS-assisted OAM secure communication system, where the legitimate transmitter (called Alice) and the legitimate receiver (called Bob) are $ N $-array UCAs to generate and receive multiple OAM signals, respectively. An eavesdrop UCA (called Eve) composed of $ N $ arrays is randomly located around Alice to intercept the legitimate OAM communications. There exists a phase difference $e^{j\frac{2\pi\l}{N}}$ between any two adjacent arrays of UCAs, where $ l \left(  -\left\lfloor \frac{N}{2}\right\rfloor+1\leq l \leq \left\lfloor \frac{N}{2}\right\rfloor\right)  $  represents the order of OAM-modes and $\left\lfloor \cdot \right\rfloor$ is the floor function.
In the RIS-assisted OAM system, a rectangular RIS with $ Q=Q_{\mathrm{y}} Q_{\mathrm{z}} $ reflecting elements is used to flexibly adjust the OAM beams direction, where $ Q_{\mathrm{y}} $ and $ Q_{\mathrm{z}} $ are the numbers of reflecting elements along the y-axis and z-axis, respectively. It is supposed that Alice and Bob are parallel and aligned.

To explain the RIS-assisted OAM system clearly, we establish a coordinate system as shown in Fig. \ref{fig: system model}.
The coordinate corresponding to the center point of UCA at Alice is denoted by $\mathbf{u}_{\mathrm{A}}=[0,0,0]^{\mathrm{T}}$, where $\left( \cdot\right)^{\mathrm{T}}$ denotes the  transpose of a matrix. The coordinates of the x-axis, y-axis, and z-axis with respect to the center point of UCA at Bob  are respectively denoted by $ x_\mathrm{B} $, $ y_\mathrm{B} $, and $ z_\mathrm{B} $. Thus, the coordinate of Bob's center point  $\mathbf{u}_{\mathrm{B}}=[x_{\mathrm{B}},y_{\mathrm{B}} ,z_{\mathrm{B}} ]^{\mathrm{T}}$. Similarly, we have the coordinate of RIS's center point  $\mathbf{u}_{\mathrm{R}}=\left[x_{\mathrm{R}}, y_{\mathrm{R}}, z_{\mathrm{R}}\right]^{\mathrm{T}}$, where $x_\mathrm{R} $, $ y_\mathrm{R} $, and $ z_\mathrm{R} $ are the coordinates along the x-axis, y-axis, and z-axis, respectively. $D$ is the distance from the center of Alice to the
center of Eve. $\theta$ denotes the included angle between x-axis and the projection of the line from the center of Alice to the center of Eve on the transmit plane. Besides, $\varphi$ denotes the included angle between z-axis and the
line from the center of Alice to the center of Eve. Thus, the center point of Eve is located at $\mathbf{u}_{\mathrm{E}}\hspace{-1mm}=\hspace{-1mm}\left[D\hspace{-0.5mm}\sin\hspace{-0.5mm}\varphi\hspace{-0.5mm}\cos\hspace{-0.5mm}\theta\hspace{-0.5mm},D\hspace{-0.5mm}\sin\hspace{-0.5mm}\varphi\hspace{-0.5mm}\sin\hspace{-0.5mm}\theta\hspace{-0.5mm},D\hspace{-0.5mm}\cos\hspace{-0.5mm}\varphi\right]^{\mathrm{T}}$. In addition, $ \alpha_{\mathrm{A}} $, $ \alpha_{\mathrm{B}} $, and $ \alpha_{\mathrm{E}} $ denote the angles between the phase angle of the first array and zero radian of Alice, Bob, and Eve, respectively. 

Hence, the coordinate of the $n$-th $ (1\leq n\leq N) $ array on Alice is given as follows: 
\begin{equation}
	\setlength{\abovedisplayskip}{3pt}
	\setlength{\belowdisplayskip}{3pt}
\begin{aligned}
 \mathbf{u}_{\mathrm{A}, n}= \mathbf{u}_{\mathrm{A}}+\left[r_{\mathrm{A}} \cos \left( \phi_{n}\right) , r_{\mathrm{A}} \sin \left( \phi_{n}\right) ,0 \right]^{\mathrm{T}},
\end{aligned}	 
\end{equation}
where $\phi_{n}=\frac{2 \pi(n-1)}{N}+\alpha_{\mathrm{A}}$ is the azimuthal angle (defined as the angular position on a plane perpendicular to the axis of propagation) for the $n$-th array on Alice, and $r_{\mathrm{A}}$ represents the radius of Alice. Similarly, the coordinates of the $m$-th $ (1\leq m\leq N) $ array on Bob and the $v$-th $ (1\leq v\leq N) $ array on Eve are represented by
\begin{equation}
	\setlength{\abovedisplayskip}{3pt}
	\setlength{\belowdisplayskip}{3pt}
\left\{\hspace{-2.5mm}
\begin{array}{lr}	
	\mathbf{u}_{\mathrm{B}, m}\hspace{-1mm}=\hspace{-1mm} \mathbf{u}_{\mathrm{B}}\hspace{-1mm}+\hspace{-1mm}\left[r_{\mathrm{B}} \cos \left(\psi_{m}\right)\hspace{-0.5mm},  r_{\mathrm{B}} \sin\left( \psi_{m}\right)\hspace{-0.5mm},0\right]^{\mathrm{T}};\\
 \mathbf{u}_{\mathrm{E}, v}=\hspace{-1mm} \mathbf{u}_{\mathrm{E}}\hspace{-1mm}+\hspace{-1mm}\mathbf{R}_{\mathrm{y}}\hspace{-0.5mm}\left(\hspace{-0.5mm}\vartheta_{ \mathrm{y}}\hspace{-0.5mm}\right)\hspace{-0.5mm}\mathbf{R}_{ \mathrm{x}}\hspace{-0.5mm}\left(\hspace{-0.5mm}\vartheta_{ \mathrm{x}}\hspace{-0.5mm}\right)\hspace{-0.5mm}\left[ r_{\mathrm{E}} \hspace{-0.5mm}\cos \hspace{-0.5mm}\left(\hspace{-0.5mm}\kappa_{v}\hspace{-0.5mm}\right)\hspace{-0.5mm},r_{\mathrm{E}} \hspace{-0.5mm}\sin \left(\hspace{-0.5mm}\kappa_{ v}\hspace{-0.5mm}\right)\hspace{-0.5mm},0\right]^{\mathrm{T}}\hspace{-2mm},
\end{array}\label{u_B_E}
\right.
\end{equation}
where $\psi_{m}=\frac{2 \pi(m-1)}{N}+\alpha_{\mathrm{B}}$ and $\kappa_{v}=\frac{2 \pi(v-1)}{N}+\alpha_{\mathrm{E}}$ are respectively the azimuthal angles for the $m$-th array on Bob and the $v$-th array on Eve,  $r_{\mathrm{B}}$ and $r_{\mathrm{E}}$ respectively represent the radii of Bob and Eve,  and
$ \mathbf{R}_{\mathrm{x}}\left(\vartheta_{\mathrm{x}}\right) $ and $ \mathbf{R}_{\mathrm{y}}\left(\vartheta_{ \mathrm{y}}\right) $  are respectively the attitude matrices  corresponding to the x-axis and y-axis. Also, $ \vartheta_{\mathrm{x}} $ and $ \vartheta_{\mathrm{y}} $  represent the rotation angles around the x-axis and y-axis for Eve, respectively.

Based on $\mathbf{u}_{\mathrm{R}}$, the coordinate of the $q$-th $ (1\leq q\leq Q) $ reflecting element on the RIS can be expressed as follows:
\begin{equation}
	\setlength{\abovedisplayskip}{3pt}
	\setlength{\belowdisplayskip}{3pt}
\begin{aligned}
\!\mathbf{u}_{\mathrm{R}, q}  \hspace{-1mm}=\hspace{-1mm} \mathbf{u}_{\mathrm{R}}\!+\!\left[\hspace{-0.5mm} 0,d_{\mathrm{y}}\hspace{-0.5mm}\left(\hspace{-0.5mm}q_{\mathrm{y}}\!+\!\frac{-1\!-\!Q_{\mathrm{y}}}{2}\hspace{-1mm}\right),d_{\mathrm{z}}\hspace{-0.5mm}\left(\hspace{-0.5mm}q_{\mathrm{z}}\!+\!\frac{-1\!-\!Q_{\mathrm{z}}}{2}\hspace{-1mm}\right)\hspace{-0.5mm}\right]^{\mathrm{T}}\!,
\end{aligned}
\end{equation}
where $q_{\mathrm{y}} \in \left\{1,2, \ldots, Q_{\mathrm{y}}\right\}$, $q_{\mathrm{z}} \in \left\{1,2, \ldots, Q_{\mathrm{z}}\right\}$, $d_{\mathrm{y}}$ and $d_{\mathrm{z}}$ are the element separation distances on the RIS along the y-axis and z-axis, respectively.

To improve the secrecy performance, we first divide all available OAM-modes into $N_\mathrm{A}$ low-order OAM-modes and $\left( N-N_\mathrm{A}\right) $ high-order OAM-modes. Due to the divergence of OAM beams and the requirement of transmission rate of legitimate receiver, zero OAM-mode is assigned to low-order OAM-modes. Next, introducing index modulation \cite{Index-Modulation}, the desired signals are transmitted by $ N_\mathrm{s}$ OAM-modes with the set $\mathcal{L}_\mathrm{s}$, where $\mathcal{L}_\mathrm{s}$ consists of zero and $\left( N_\mathrm{s}-1\right)  $ non-zero OAM-modes randomly chosen from $\left( N_\mathrm{A}-1\right) $ low-order OAM-modes. Similarly, AN is transmitted by $ N_\mathrm{\ddot{z}}$ OAM-modes in the set $\mathcal{L_\mathrm{\ddot{z}}}$, where the $ N_\mathrm{\ddot{z}}$ OAM-modes are randomly chosen from $\left( N-N_\mathrm{A}\right) $ high-order OAM-modes. 
Each combination of $\mathcal{L}_\mathrm{s}$ and $\mathcal{L_\mathrm{\ddot{z}}}$ corresponds to the specific index information for the OAM-modes used to transmit the desired signals and AN. For example, we successively pre-set the sets $\mathcal{L}_\mathrm{s}=\left\lbrace 0,+1,-1\right\rbrace $ and $\left\lbrace 0,+1,-2\right\rbrace $ for two consecutive desired signal transmissions.
Therefore, the number of combinations of $\mathcal{L}_\mathrm{s}$ and $\mathcal{L}_\mathrm{\ddot{z}}$ is $K\!=\! 2^{\left\lfloor\log_{2}\left[  \!\left(\!\!\!\!
	\begin{tiny}
		\begin{array}{c}
			N_\mathrm{A}-1 \\
			N_\mathrm{s}-1
		\end{array}
	\end{tiny} \!\!\!\!\right) \left(\!\!\!\!
	\begin{tiny}
		\begin{array}{c}
			N\!-\!N_\mathrm{A} \\
			N_\mathrm{\ddot{z}}
		\end{array}
	\end{tiny} \!\!\!\!\right)\right] \right\rfloor }$, where $\left(\cdot \right)$ denotes the binomial coefficient. $\mathcal{L}_\mathrm{s} $ and $\mathcal{L}_\mathrm{\ddot{z}} $ are known by the legitimate transceivers. By introducing index modulation, the spectrum efficiency of our proposed scheme is significantly increased at the low hardware cost of radio frequency chains. Meanwhile, the randomness of OAM-modes selected by our proposed scheme could contribute to preventing information leakage. 

Next, we denote $\tilde{\mathbf{s}}=\mathbf{s}+\mathbf{z} $  the initial signal vector corresponding to all OAM-modes, where $\mathbf{s} \in \mathbb{C}^{N \times 1}$ denotes the desired signal vector and $\mathbf{z} \in \mathbb{C}^{N \times 1}$ denotes the AN vector.  In $\mathbf{s}$, all other entries are 0 except
the entries selected to transmit the desired signals with the non-zero entry $\mathbb{E}\left\lbrace \left\vert s_{l^\mathrm{s}_{\bar{n}}} \right\vert^2\right\rbrace  =p_{l^\mathrm{s}_{\bar{n}}} $, where  $ l^\mathrm{s}_{\bar{n}}$ is the $ \bar{n}$-th $ \left( 1\leq \bar{n} \leq N_\mathrm{s}\right) $ OAM-mode used to transmit the desired signals in $\mathcal{L}_\mathrm{s}  $, $ s_{l^\mathrm{s}_{\bar{n}}}$ is the transmit symbol, $ p_{l^\mathrm{s}_{\bar{n}}}$ is the allocated power  to the  OAM-mode $l^\mathrm{s}_{\bar{n}}$, and $\vert \cdot\vert $ is the absolute value of a scalar. We denote by $\rho$ $\left(0<\rho \leq 1 \right) $ the ratio of the transmit power for the desired signals to the total transmit power $P_\mathrm{T}$. Similarly, in $\mathbf{z}$ all other entries are 0 except that
the entries selected to transmit AN and the non-zero entry $ z_{l^\mathrm{z}_{\ddot{n}}}\sim \mathcal{C N}\left(0, \sigma_\mathrm{\ddot{z}}^{2}\right)$ with variance $ \sigma_\mathrm{\ddot{z}}^{2}$
, where $ l^\mathrm{z}_{\ddot{n}}$ is the $ \ddot{n}$-th $ \left( 1\leq \ddot{n} \leq N_\mathrm{\ddot{z}}\right) $ OAM-mode used to transmit AN in $\mathcal{L}_\mathrm{\ddot{z}}  $. For simplicity, OAM-modes used to transmit AN are allocated to equal power, that is $\sigma_\mathrm{\ddot{z}}^2=\frac{\left( 1-\rho\right)P_\mathrm{T} }{N_\mathrm{\ddot{z}}}$. 

To obtain the transmit signal $\mathbf{x} $, the initial signal $\tilde{\mathbf{s}}$ is first modulated into the antenna excitations with equal amplitudes and linearly increasing phases of $2\pi l/N$ for different OAM-modes. This modulation is equivalent to performing the inverse discrete Fourier inverse transform (IDFT) on $\tilde{\mathbf{s}}$. 
Hence, the transmit signal $\mathbf{x} $ for the developed system is given as follows:
\begin{equation}
	\setlength{\abovedisplayskip}{3pt}
	\setlength{\belowdisplayskip}{3pt}
\mathbf{x} = \mathbf{F } \tilde{\mathbf{s}},\label{con:transmit signal}
\end{equation}
where $\mathbf{F}\hspace{-1mm}=\hspace{-1mm}\left[\mathbf{f}_{0}, \ldots, \mathbf{f}_{N-1}\right]  \mathbb{C}^{N \times N}$ is the IDFT matrix with $\mathbf{f}_{l}\hspace{-1mm}=\hspace{-1mm}\frac{1}{\sqrt{N}}\hspace{-1mm}\left[\hspace{-0.5mm} e^{j l  \phi_{1} }\hspace{-1mm}, \ldots,\hspace{-0.5mm} e^{j l \phi_{N}}\hspace{-1mm}\right]^\mathrm{T}$. 

Then, the transmit signals are reflected by the RIS to adjust OAM beams received by Bob and Eve. We $ $denote by $\mathbf{\Theta}=\operatorname{diag}\left(\boldsymbol{\theta}^\mathrm{H}\right)$ the diagonal phase shift matrix of RIS with $\boldsymbol{\theta}=\left[\theta_{1}, \ldots, \theta_{Q}\right]^\mathrm{T}$, where $\theta_{q} = e^{j\hat{\theta}_{q}}$ denote the  phase shift of the $q$-th reflecting element with $\hat{\theta}_{q} \in \left[ 0,2\pi\right] $ and the unit modulus $ \left|\theta_{q}\right|  = 1$, $ \forall q \in \left\lbrace 1,\dots, Q\right\rbrace$, and $(\cdot)^\mathrm{H}$ denotes the conjugate transpose of a matrix. Thus, we respectively have the received signals, denoted by $ \mathbf{y}_{\mathrm{B}} $ and $ \mathbf{y}_{\mathrm{E}} $, at Bob and Eve as follows:
\begin{equation}
	\setlength{\abovedisplayskip}{3pt}
	\setlength{\belowdisplayskip}{3pt}
	\left\{
	\begin{array}{lr}
\mathbf{y}_{\mathrm{B}} =\left(\mathbf{H}_{\mathrm{AB}}+ \mathbf{H}_{\mathrm{RB}} \boldsymbol{\Theta} \mathbf{H}_{\mathrm{AR}}\right) \mathbf{x}+\mathbf{n}_{\mathrm{B}};\\
\mathbf{y}_{\mathrm{E}} = \left(\mathbf{H}_{\mathrm{AE}}+\mathbf{H}_{\mathrm{RE}} \mathbf{\Theta} \mathbf{H}_{\mathrm{AR}}\right) \mathbf{x}+\mathbf{n}_{\mathrm{E}},
	\end{array}\label{con:y_B_E}
\right.
\end{equation}
where $\mathbf{n}_\mathrm{B} \sim \mathcal{C N}\left(0, \sigma_\mathrm{B}^{2} \mathbf{I}_{N}\right)$ is the received Gaussian noise at Bob with variance $ \sigma_\mathrm{B}^{2} $, $\mathbf{n}_\mathrm{E} \sim \mathcal{C N}\left(0, \sigma_\mathrm{E}^{2} \mathbf{I}_{N}\right)$ is the received Gaussian noise at Eve with variance $ \sigma_\mathrm{E}^{2} $, and $ \mathbf{I}_{N} $ denotes the $ N\times N $ identity matrix. Also, $\mathbf{H}_{\mathrm{AB}}\in \mathbb{C}^{N \times N} $, $\mathbf{H}_{\mathrm{AR}} \in \mathbb{C}^{Q \times N}$, $\mathbf{H}_{\mathrm{RB}} \in \mathbb{C}^{N \times Q}$, $\mathbf{H}_{\mathrm{AE}} \in \mathbb{C}^{N \times N}$, and $\mathbf{H}_{\mathrm{RE}} \in \mathbb{C}^{N \times Q}$ denote the channels from Alice to Bob, Alice to RIS, RIS to Bob, Alice to Eve, and RIS to Eve, respectively. Defining $\tau \in \left\lbrace {\mathrm{A,R}}\right\rbrace  $, $\kappa \in \left\lbrace {\mathrm{B,R,E}}\right\rbrace  $, $\delta \in \left\lbrace {m,q,v}\right\rbrace  $, and $\chi \in \left\lbrace {n,q}\right\rbrace  $, we have the LoS channels from the $ \chi$-th array of the end $ \tau$ (Alice or RIS) to the $ \delta$-th array of the end $\kappa$ (Bob, RIS, or Eve), denoted by $ h_{\tau \kappa, \delta \chi}$, as follows: 
	\begin{equation}
	\setlength{\abovedisplayskip}{3pt}
		\setlength{\belowdisplayskip}{3pt}
		h_{\tau \kappa, \delta \chi}  = \frac{\beta \lambda}{4 \pi \Vert \mathbf{u}_{\kappa,\delta }-\mathbf{u}_{\tau, \chi}\Vert} e^{-j \frac{2 \pi }{\lambda} \Vert \mathbf{u}_{\kappa,\delta }-\mathbf{u}_{\tau, \chi}\Vert},\label{channels}
	\end{equation}
where $\beta$ represents the attenuation, $\lambda$ is the wavelength, and $ \Vert \cdot\Vert $ is the Euclidean norm of a vector. 
Thus, the entries of $h_{\mathrm{AB},mn}$, $h_{\mathrm{AR},qn}$, $h_{\mathrm{RB},mq}$, $h_{\mathrm{AE},vn}$, and $h_{\mathrm{RE},vq}$ corresponding to $\mathbf{H}_{\mathrm{AB}}$, $\mathbf{H}_{\mathrm{AR}}$, $\mathbf{H}_{\mathrm{RB}}$, $\mathbf{H}_{\mathrm{AE}}$, and $\mathbf{H}_{\mathrm{RE}}$, respectively, are given by Eq.~\eqref{channels}.


Based on Eqs. (\ref{con:transmit signal}) and (\ref{con:y_B_E}), the decomposed signals, denoted by $\tilde{\mathbf{y}}_{\mathrm{B}} \in \mathbb{C}^{N \times 1}$, at Bob with the discrete Fourier transform (DFT) are derived as follows:
\begin{equation}
	\setlength{\abovedisplayskip}{3pt}
	\setlength{\belowdisplayskip}{3pt}
\begin{aligned}
\tilde{\mathbf{y}}_{\mathrm{B}}= \mathbf{F}^\mathrm{H}\left( \mathbf{H}_{\mathrm{AB}}+ \mathbf{H}_{\mathrm{RB}} \mathbf{\Theta} \mathbf{H}_{\mathrm{AR}}\right)  \mathbf{F }\mathbf{s}+\tilde{\mathbf{n}}_{\mathrm{B}},\label{con:DFT-yB}
\end{aligned}
\end{equation}
where $\tilde{\mathbf{n}}_{\mathrm{B}}=\mathbf{F}^\mathrm{H} \mathbf{n}_{\mathrm{B}}$ follows the Gaussian distribution with variance $ \sigma_{\mathrm{B}}^{2} $.
Since the orthogonality of OAM beams reflected by RIS is destroyed, there exists inter-mode interference after OAM signal decomposition. Also, AN can be easily eliminated because AN is pre-known by Alice and Bob. Thus, the signal-to-interference-plus-noise ratio (SINR), denoted by $ \gamma_{\mathrm{B}, l^\mathrm{s}_{\bar{n}}} $, for Bob with respect to the OAM-mode $l^\mathrm{s}_{\bar{n}}$ is derived as follows:
\begin{equation}
	\setlength{\abovedisplayskip}{3pt}
	\setlength{\belowdisplayskip}{3pt}
\begin{aligned}
		\gamma_{\mathrm{B}, l^\mathrm{s}_{\bar{n}}} = \frac{p_{l^\mathrm{s}_{\bar{n}}}\left| a_{l^\mathrm{s}_{\bar{n}},l^\mathrm{s}_{\bar{n}}} \right|^{2}}{\sum\limits_{ k \neq l^\mathrm{s}_{\bar{n}}, k\in \mathcal{L}_\mathrm{s}} \!\!\!\!p_{k}\left| a_{l^\mathrm{s}_{\bar{n}},k} \right|^{2}+1},
	\label{con:gamma_B}
\end{aligned}
\setlength{\belowdisplayskip}{3pt}
\end{equation}
where $a_{l^\mathrm{s}_{\bar{n}},k}=\sigma_{\mathrm{B}}^{-1} \mathbf{f}_{l^\mathrm{s}_{\bar{n}}}^\mathrm{H} \left( \mathbf{H}_{\mathrm{AB}}+ \mathbf{H}_{\mathrm{RB}} \mathbf{ \Theta } \mathbf{H}_{\mathrm{AR}}\right)  \mathbf{f}_{k}$ with $k\in \mathcal{L}_\mathrm{s}$.
Eve is unaware that the legitimate signals sent by Alice are carried with OAM.  Also, existing Eve mainly eavesdrops on legitimate signals with traditional MIMO. Hence, the SINR, denoted by $ \gamma_{\mathrm{E}, l^\mathrm{s}_{\bar{n}}} $,  with respect to the OAM-mode $l^\mathrm{s}_{\bar{n}}$ for Eve can be calculated as follows:
\begin{equation}
	\setlength{\abovedisplayskip}{3pt}
	\setlength{\belowdisplayskip}{3pt}
\begin{aligned}
		\gamma_{\mathrm{E}, l^\mathrm{s}_{\bar{n}}} = \frac{p_{l^\mathrm{s}_{\bar{n}}}\left| b_{l^\mathrm{s}_{\bar{n}},l^\mathrm{s}_{\bar{n}}} \right|^{2}}{\sum\limits_{k \neq l^\mathrm{s}_{\bar{n}},k\in \mathcal{L}_\mathrm{s}}\!\!\!\! p_{k}\left| b_{l^\mathrm{s}_{\bar{n}},k} \right|^{2}+\sigma_\mathrm{\ddot{z}}^2\sum\limits_{l^\mathrm{z}_{\ddot{n}} \in \mathcal{L}_\mathrm{\ddot{z}}}\!\!\!\left|b_{l^\mathrm{s}_{\bar{n}},l^\mathrm{z}_{\ddot{n}}} \right|^{2}+1},
\label{con:gamma_E}
\end{aligned}
\end{equation}
where  $b_{l^\mathrm{s}_{\bar{n}},\ddot{k}}=\sigma_{\mathrm{E}}^{-1}  \left(\mathbf{h}_{\mathrm{AE},l^\mathrm{s}_{\bar{n}}}+\mathbf{h}_{\mathrm{RE},l^\mathrm{s}_{\bar{n}}} \mathbf{\Theta} \mathbf{H}_{\mathrm{AR}}\right) \mathbf{f}_{\ddot{k}}$ with $\ddot{k}\in \left\lbrace k,l^\mathrm{z}_{\ddot{n}}\right\rbrace $, $\mathbf{h}_{\mathrm{AE},l^\mathrm{s}_{\bar{n}}}$ and $\mathbf{h}_{\mathrm{RE},l^\mathrm{s}_{\bar{n}}}$ represent the $l^\mathrm{s}_{\bar{n}}$-th row of $\mathbf{H}_{\mathrm{AE}}$ and $\mathbf{H}_{\mathrm{RE}}$, respectively. 
Hence, the achievable rate at Bob is calculated by 
	\begin{equation}
		\setlength{\abovedisplayskip}{3pt}
		\setlength{\belowdisplayskip}{3pt}
		R_\mathrm{B}=\tilde{R}_\mathrm{B}+\log_{2}K,
	\end{equation}
	where $ \tilde{R}_\mathrm{B}=\sum\limits_{l^\mathrm{s}_{\bar{n}}\in \mathcal{L}_\mathrm{s}} \log _{2}\left(1+\gamma_{\mathrm{B}, l^\mathrm{s}_{\bar{n}}}\right)$ is the signal information and $\log_{2}K$ is the index information [4]. 
	With given $N$, $N_{\mathrm{A}}$, $N_\mathrm{s}$, and $ N_{\ddot{\mathrm{z}}}$, $\log_{2}K$ is a constant.
Also, the achievable rate at  Eve is  $R_{\mathrm{E}}=\sum\limits_{l^\mathrm{s}_{\bar{n}}\in \mathcal{L}_\mathrm{s}} \log _{2}\left(1+\gamma_{\mathrm{E}, l^\mathrm{s}_{\bar{n}}}\right)$. 

\vspace{-0.1cm}
\subsection{Problem Formulation}
\vspace{-0.1cm}
Based on the above analysis, the secrecy rate can be calculated by $ \left( R_{\mathrm{B}}-R_{\mathrm{E}}\right)  $. Aiming at maximizing the secrecy rate , we jointly optimize the transmit power allocation $\mathbf{p}_\mathrm{s}=[p_{l^\mathrm{s}_{1}},\ldots,p_{l^\mathrm{s}_{N_\mathrm{s}}} ]^{\mathrm{T}}  $ for the desired signals and the RIS’s phase shift $\boldsymbol{\theta}$ subject to the constraints of total transmit power of desired signals and unit modulus of phase shifts. Therefore, the optimization problem can be formulated as follows:
\begin{subequations}
	\setlength{\abovedisplayskip}{3pt}
	\setlength{\belowdisplayskip}{3pt}	
\begin{align}	
		\textbf{P1}: \; \max _{\mathbf{p}_\mathrm{s}, \boldsymbol{\theta}} \;  & R_{\mathrm{B}}-R_{\mathrm{E}}\\
		\mathrm{s.t}.\quad\,\!\!  & \sum\limits_{l^\mathrm{s}_{\bar{n}}\in \mathcal{L}_\mathrm{s}} p_{l^\mathrm{s}_{\bar{n}}} \leq \rho P_\mathrm{T},\    p_{l^\mathrm{s}_{\bar{n}}} \geq p_\mathrm{th}, \forall l^\mathrm{s}_{\bar{n}} \in \mathcal{L}_\mathrm{s};\\
		&\left|\theta_{q}\right|  = 1, \forall q \in \left\lbrace 1,\dots, Q\right\rbrace,
\end{align}	
\end{subequations}
where $ p_\mathrm{th}$ $\left( p_\mathrm{th}>0\right)  $ is the allocated power threshold.  Since problem $ \textbf{P1}  $ is a non-convex problem with coupled optimization variables, it is very difficult to obtain the optimal solution. Also, the non-convex unit modulus constraints aggravate the difficulty. Therefore, it is required to find an effective optimization algorithm to achieve the maximum secrecy rate.

\vspace{-0.1cm}
\section{Proposed Algorithm}\label{sec:Solution 3}
\vspace{-0.1cm}
In this section, we develop the RMCG-AO algorithm to tackle problem $ \textbf{P1}  $. Specifically, we first decompose $ \textbf{P1}  $ into two subproblems: 1) optimizing $ \mathbf{p}_\mathrm{s} $ with given $ \boldsymbol{\theta} $; 2) optimizing $ \boldsymbol{\theta} $ with given $ \mathbf{p}_\mathrm{s} $; and then alternately solve the subproblems.

\vspace{-0.2cm}
\subsection{Optimize $\mathbf{p}_\mathrm{s}$ with given $\boldsymbol{\theta}$}
\vspace{-0.2cm}
For given $\boldsymbol{\theta}$, we define $\mathbf{a}_{l^\mathrm{s}_{\bar{n}}}\!\!=\!\!\left[\!\vert a_{l^\mathrm{s}_{\bar{n}}\!, l^\mathrm{s}_{1}}\vert^{2}\!\!, \ldots,\!\left\vert a_{l^\mathrm{s}_{\bar{n}}, l^\mathrm{s}_{N_\mathrm{s}}}\right\vert^{2}\right]^\mathrm{T}$ 
and  $\mathbf{a}_{-l^\mathrm{s}_{\bar{n}}}\!\!=\!\!\left[\!\left| a_{l^\mathrm{s}_{\bar{n}}, l^\mathrm{s}_{1}}\!\right|^{2}\!\!, \ldots,\!\left|a_{l^\mathrm{s}_{\bar{n}}, l^\mathrm{s}_{\bar{n}}-1}\!\right|^{2}\!\!,0,\!\left| a_{l^\mathrm{s}_{\bar{n}}, l^\mathrm{s}_{\bar{n}}+1}\!\right|^{2}\!\!,\ldots,\!\left| a_{l^\mathrm{s}_{n_\mathrm{s}}, l^\mathrm{s}_{N_\mathrm{s}}}\!\right|^{2}\right]^\mathrm{T}$. Thus, $  \tilde{R}_{\mathrm{B}}$ can be re-written as follows:
\begin{equation}
	\setlength{\abovedisplayskip}{3pt}
	\setlength{\belowdisplayskip}{-3pt}
\tilde{R}_{\mathrm{B}}\!= \!\sum_{l^\mathrm{s}_{\bar{n}}\in \mathcal{L}_\mathrm{s}} \left[ \log _{2}\left(1+\mathbf{a}_{l^\mathrm{s}_{\bar{n}}}^\mathrm{T} \mathbf{p}_\mathrm{s}\right)-\log _{2}\left(1+\mathbf{a}_{-l^\mathrm{s}_{\bar{n}}}^\mathrm{T} \mathbf{p}_\mathrm{s}\right)\right] .
\end{equation}
Similarly, we denote $\mathbf{b}_{l^\mathrm{s}_{\bar{n}}}=\left[\left| b_{l^\mathrm{s}_{\bar{n}}, l^\mathrm{s}_{1}}\right|^{2}, \ldots,\left| b_{l^\mathrm{s}_{\bar{n}}, l^\mathrm{s}_{N_\mathrm{s}}}\right|^{2}\right]^\mathrm{T}$ and $\mathbf{b}_{-l^\mathrm{s}_{\bar{n}}}\!\!=\!\!\left[\!\left| b_{l^\mathrm{s}_{\bar{n}}, l^\mathrm{s}_{1}}\!\right|^{2}\!\!, \ldots,\!\left|b_{l^\mathrm{s}_{\bar{n}}, l^\mathrm{s}_{\bar{n}}-1}\!\right|^{2}\!\!,0,\!\left| b_{l^\mathrm{s}_{\bar{n}}, l^\mathrm{s}_{\bar{n}}+1}\!\right|^{2}\!\!,\ldots,\!\left| b_{l^\mathrm{s}_{n_\mathrm{s}}, l^\mathrm{s}_{N_\mathrm{s}}}\!\right|^{2}\right]^\mathrm{T}$. Thus, $  R_{\mathrm{E}}$ can be re-expressed as follows:
\begin{equation}
	\setlength{\abovedisplayskip}{3pt}
	\setlength{\belowdisplayskip}{3pt}
\begin{aligned}
R_{\mathrm{E}}\!= \!\!\!\sum_{l^\mathrm{s}_{\bar{n}}\in \mathcal{L}_\mathrm{s}} \!\!\!\!\left[ \log _{2}\!\left(\!1\!+\!\mathbf{b}_{l^\mathrm{s}_{\bar{n}}}^\mathrm{T} \mathbf{p}_\mathrm{s}\!+\! c\right)\!-\!\log _{2}\!\left(1\!+\!\mathbf{b}_{-l^\mathrm{s}_{\bar{n}}}^\mathrm{T} \mathbf{p}_\mathrm{s}\!+\!c\right)\!\right],
\end{aligned}
\end{equation}
where $c=\sigma_\mathrm{\ddot{z}}^2\!\!\sum\limits_{l^\mathrm{z}_{\ddot{n}} \in \mathcal{L}_\mathrm{\ddot{z}}}\!\!\!\left|b_{l^\mathrm{s}_{\bar{n}},l^\mathrm{z}_{\ddot{n}}} \right|^{2}$.
Consequently, problem $ \textbf{P1} $ can be re-formulated as follows:
\begin{subequations}
	\setlength{\abovedisplayskip}{3pt}	
	\setlength{\belowdisplayskip}{5pt}
	\begin{align}	
		\textbf{P2}: \;\  \max _{\mathbf{p}_\mathrm{s} }\; &\tilde{R}_{\mathrm{B}}-R_{\mathrm{E}} \\
		\mathrm{s.t}.\;\, & \sum\limits_{l^\mathrm{s}_{\bar{n}}\in \mathcal{L}_\mathrm{s}} p_{l^\mathrm{s}_{\bar{n}}} \leq \rho P_\mathrm{T},\  p_{l^\mathrm{s}_{\bar{n}}} \geq p_\mathrm{th}, \forall l^\mathrm{s}_{\bar{n}}\in \mathcal{L}_\mathrm{s}.
	\end{align}	
\end{subequations}
Note that $\log_{2}K$  can be omitted in \textbf{P2}. To solve the non-convex problem $ \textbf{P2} $, \textbf{Lemma 1} is applied \cite{Transmit_Solutions_for_MIMO_AO}.
\begin{figure*}[hbp]
	\setcounter{equation}{24}
	\vspace*{-16pt}
	\hrulefill
	\vspace*{-5pt}
	\begin{equation}
		\begin{scriptsize}
			\begin{aligned}
				&\nabla_{\boldsymbol{\theta}_{i}} f\!=\hspace{-1mm}\sum\limits_{l^\mathrm{s}_{\bar{n}}\in \mathcal{L}_\mathrm{s}} \hspace{-1mm}\dfrac{2}{\ln2}  \left(-\dfrac{\sum\limits_{k\in \mathcal{L}_\mathrm{s}}\hspace{-1mm} p_{k}\hspace{-1mm}\left( \boldsymbol{\mu}_{l^\mathrm{s}_{\bar{n}}\!,k}\boldsymbol{\mu}_{l^\mathrm{s}_{\bar{n}}\!,k}^\mathrm{H}\boldsymbol{\theta}_{i}+\boldsymbol{\mu}_{l^\mathrm{s}_{\bar{n}}\!,k}\omega^{*} _{l^\mathrm{s}_{\bar{n}}\!,k}\right) }{\sum\limits_{ k\in \mathcal{L}_\mathrm{s}}\hspace{-1mm} p_{k}\hspace{-0.5mm}\left|\omega _{l^\mathrm{s}_{\bar{n}}\!,k}\hspace{-0.5mm}+\hspace{-0.5mm} \boldsymbol{\theta}_{i}^\mathrm{H} \boldsymbol{\mu}_{l^\mathrm{s}_{\bar{n}},k} \right|^{2}+\sigma_{\mathrm{B}}^{2}}+\dfrac{\sum\limits_{k\neq l^\mathrm{s}_{\bar{n}}\!,k\in \mathcal{L}_\mathrm{s}} \hspace{-4mm}p_{k}\hspace{-0.5mm}\left( \boldsymbol{\mu}_{l^\mathrm{s}_{\bar{n}}\!,k}\boldsymbol{\mu}_{l^\mathrm{s}_{\bar{n}}\!,k}^\mathrm{H}\boldsymbol{\theta}_{i}+\boldsymbol{\mu}_{l^\mathrm{s}_{\bar{n}}\!,k}\omega^{*} _{l^\mathrm{s}_{\bar{n}}\!,k}\right)}{\sum\limits_{k\neq l^\mathrm{s}_{\bar{n}},k\in \mathcal{L}_\mathrm{s}}\hspace{-1mm} p_{k}\hspace{-0.5mm}\left|\omega _{l^\mathrm{s}_{\bar{n}}\!,k}\hspace{-1mm}+\hspace{-0.5mm} \boldsymbol{\theta}_{i}^\mathrm{H} \boldsymbol{\mu}_{l^\mathrm{s}_{\bar{n}},k} \right|^{2}+\sigma_{\mathrm{B}}^{2}} \right.\\
				& \left.+\dfrac{\sum\limits_{k\in \mathcal{L}_\mathrm{s}} \hspace{-2mm}p_{k}\hspace{-1mm}\left(\hspace{-0.5mm}\boldsymbol{\eta}_{l^\mathrm{s}_{\bar{n}}\!,k}\boldsymbol{\eta}_{l^\mathrm{s}_{\bar{n}}\!,k}^\mathrm{H}\boldsymbol{\theta}_{i}\hspace{-1mm}+\hspace{-0.5mm}\boldsymbol{\eta}_{l^\mathrm{s}_{\bar{n}}\!,k}\zeta_{l^\mathrm{s}_{\bar{n}}\!,k}^{*}\hspace{-0.5mm}\right)\hspace{-1mm}+\hspace{-0.5mm}\sigma_\mathrm{\ddot{z}}^2\hspace{-2mm}\sum\limits_{l^\mathrm{z}_{\ddot{n}}\!\in \mathcal{L}_\mathrm{\ddot{z}}}\hspace{-3mm}\left(\hspace{-0.5mm}\boldsymbol{\eta}_{l^\mathrm{s}_{\bar{n}}\!,l^\mathrm{z}_{\ddot{n}}}\boldsymbol{\eta}_{l^\mathrm{s}_{\bar{n}}\!,l^\mathrm{z}_{\ddot{n}}}^\mathrm{H}\boldsymbol{\theta}_{i}\hspace{-0.5mm}+\hspace{-0.5mm}\boldsymbol{\eta}_{l^\mathrm{s}_{\bar{n}}\!,l^\mathrm{z}_{\ddot{n}}}\zeta_{l^\mathrm{s}_{\bar{n}}\!,l^\mathrm{z}_{\ddot{n}}}^{*}\hspace{-0.5mm} \right) }{\sum\limits_{k\in \mathcal{L}_\mathrm{s}} \hspace{-1.5mm}p_{k}\hspace{-0.5mm}\left|  \zeta_{l^\mathrm{s}_{\bar{n}},k}\hspace{-0.5mm}+\hspace{-0.5mm} \boldsymbol{\theta}_{i}^\mathrm{H} \boldsymbol{\eta}_{l^\mathrm{s}_{\bar{n}},k}\hspace{-0.5mm} \right|^{2}\hspace{-0.5mm}+\sigma_\mathrm{\ddot{z}}^2\hspace{-1mm}\sum\limits_{ l^\mathrm{z}_{\ddot{n}} \!\in \mathcal{L}_\mathrm{\ddot{z}}}\!\! \left|  \zeta_{l^\mathrm{s}_{\bar{n}}\!,l^\mathrm{z}_{\ddot{n}}}+ \boldsymbol{\theta}^\mathrm{H} \boldsymbol{\eta}_{l^\mathrm{s}_{\bar{n}}\!,l^\mathrm{z}_{\ddot{n}}}\! \right|^{2}\!+\!\sigma_{\mathrm{E}}^{2}}\!-\!\dfrac{\sum\limits_{k\neq l^\mathrm{s}_{\bar{n}}\!, k\in \mathcal{L}_\mathrm{s}} \hspace{-5mm}p_{k}\hspace{-1mm}\left(\hspace{-1mm}\boldsymbol{\eta}_{l^\mathrm{s}_{\bar{n}}\!,k}\boldsymbol{\eta}_{l^\mathrm{s}_{\bar{n}}\!,k}^\mathrm{H}\boldsymbol{\theta}_{i}\hspace{-0.5mm}+\hspace{-0.5mm}\boldsymbol{\eta}_{l^\mathrm{s}_{\bar{n}}\!,k}\zeta_{l^\mathrm{s}_{\bar{n}}\!,k}^{*}\hspace{-0.5mm}\right)\hspace{-0.5mm}+\hspace{-0.5mm}\sigma_\mathrm{\ddot{z}}^2\hspace{-1mm}\sum\limits_{l^\mathrm{z}_{\ddot{n}}\in \mathcal{L}_\mathrm{\ddot{z}}}\hspace{-2.5mm}\left(\hspace{-0.5mm}\boldsymbol{\eta}_{l^\mathrm{s}_{\bar{n}}\!,l^\mathrm{z}_{\ddot{n}}}\boldsymbol{\eta}_{l^\mathrm{s}_{\bar{n}}\!,l^\mathrm{z}_{\ddot{n}}}^\mathrm{H}\boldsymbol{\theta}_{i}\hspace{-0.5mm}+\hspace{-0.5mm}\boldsymbol{\eta}_{l^\mathrm{s}_{\bar{n}}\!,l^\mathrm{z}_{\ddot{n}}}\zeta_{l^\mathrm{s}_{\bar{n}}\!,l^\mathrm{z}_{\ddot{n}}}^{*} \hspace{-0.5mm}\right) }{\sum\limits_{k\neq l^\mathrm{s}_{\bar{n}},k\in \mathcal{L}_\mathrm{s}} \hspace{-4mm}p_{k}\!\left|  \zeta_{l^\mathrm{s}_{\bar{n}},k}\hspace{-0.5mm}+\hspace{-0.5mm} \boldsymbol{\theta}_{i}^\mathrm{H} \boldsymbol{\eta}_{l^\mathrm{s}_{\bar{n}},k} \right|^{2}\hspace{-0.5mm}+\hspace{-0.5mm}\sigma_\mathrm{\ddot{z}}^2\!\hspace{-0.5mm}\sum\limits_{ l^\mathrm{z}_{\ddot{n}} \!\in \mathcal{L}_\mathrm{\ddot{z}}}\!\! \hspace{-1mm}\left|  \zeta_{l^\mathrm{s}_{\bar{n}}\!,l^\mathrm{z}_{\ddot{n}}}+ \boldsymbol{\theta}^\mathrm{H} \boldsymbol{\eta}_{l^\mathrm{s}_{\bar{n}}\!,l^\mathrm{z}_{\ddot{n}}}\hspace{-0.5mm} \right|^{2}\hspace{-0.5mm}+\hspace{-0.5mm}\sigma_{\mathrm{E}}^{2}}\hspace{-1mm}\right)\hspace{-1mm}.\label{con:Euclidean gradient}
			\end{aligned}
		\end{scriptsize}
	\end{equation}
	
	\vspace*{-25pt}
\end{figure*}

\textbf{Lemma 1}:  For any given $x>0$, the maximum of the function $\varphi(t)=-t x+\ln t+1$ with the auxiliary variable $t$ is given by
\vspace{-0.2cm}
\setcounter{equation}{14}
\begin{equation}
-\ln x=\max _{t>0} \varphi(t).\label{lemma1}
\vspace{-0.1cm}
\end{equation}
The optimal solution, denoted by $t^{\mathrm{opt}}$, in Eq. (\ref{lemma1}) is $t^{\mathrm{opt}}\hspace{-1mm}=\hspace{-1mm}1/x$. 

Based on \textbf{Lemma 1},  setting $ x=1+\mathbf{a}_{-l^\mathrm{s}_{\bar{n}}}^\mathrm{T} \mathbf{p}_\mathrm{s} $ and $ t=t_{\mathrm{B},l^\mathrm{s}_{\bar{n}}} $, we have $  \tilde{R}_{\mathrm{B}}\ln2$ as follows:
\begin{equation}
	\setlength{\abovedisplayskip}{3pt}
	\setlength{\belowdisplayskip}{3pt}
\begin{aligned}
\tilde{R}_{\mathrm{B}}\ln2= \sum\limits_{l^\mathrm{s}_{\bar{n}}\in \mathcal{L}_\mathrm{s}} \max _{t_{\mathrm{B},l^\mathrm{s}_{\bar{n}}}>0} \varphi_{\mathrm{B},l^\mathrm{s}_{\bar{n}}}(\mathbf{p}_\mathrm{s},t_{\mathrm{B},l^\mathrm{s}_{\bar{n}}}),\label{con:Rbln2}
\end{aligned}
\end{equation}
where
$
\varphi_{\mathrm{B},l^\mathrm{s}_{\bar{n}}}(\mathbf{p}_\mathrm{s},t_{\mathrm{B},l^\mathrm{s}_{\bar{n}}})\!\!=\!\!\ln\left (\!1\!+\!\mathbf{a}_{l^\mathrm{s}_{\bar{n}}}^\mathrm{T} \mathbf{p}_\mathrm{s}\!\right)\!-\!t_{\mathrm{B},l^\mathrm{s}_{\bar{n}}}\left (\!1\!+\!\mathbf{a}_{-l^\mathrm{s}_{\bar{n}}}^\mathrm{T} \mathbf{p}_\mathrm{s}\!\right)+\ln t_{\mathrm{B},l^\mathrm{s}_{\bar{n}}}+1$.
Similarly, $R_{\mathrm{E}}\ln2$ can be calculated as follows:
\begin{equation}
	\setlength{\abovedisplayskip}{3pt}
	\setlength{\belowdisplayskip}{3pt}
\begin{aligned}
R_{\mathrm{E}}\ln2= \sum\limits_{l^\mathrm{s}_{\bar{n}}\in \mathcal{L}_\mathrm{s}} \min _{t_{\mathrm{E},l^\mathrm{s}_{\bar{n}}}>0} \varphi_{\mathrm{E},l^\mathrm{s}_{\bar{n}}}(\mathbf{p}_\mathrm{s},t_{\mathrm{E},l^\mathrm{s}_{\bar{n}}}),\label{con:Reln2}
\end{aligned}
\end{equation}
where
$
\varphi_{\mathrm{E},l^\mathrm{s}_{\bar{n}}}(\mathbf{p}_\mathrm{s},t_{\mathrm{E},l^\mathrm{s}_{\bar{n}}})\!=\! t_{\mathrm{E},l^\mathrm{s}_{\bar{n}}}\!\left (\!1+\!\mathbf{b}_{l^\mathrm{s}_{\bar{n}}}^\mathrm{T} \mathbf{p}_\mathrm{s}\!+\!c\!\right)\!-\!
\ln\!\left (\!1\!+\!\mathbf{b}_{-l^\mathrm{s}_{\bar{n}}}^\mathrm{T} \mathbf{p}_\mathrm{s}\notag \right.
\\
\left.+c\right)-\ln t_{\mathrm{E},l}-1$.
Hence, problem $ \textbf{P2} $ can be transformed into problem $ \textbf{P3} $ as follows:
\begin{subequations}
	\setlength{\abovedisplayskip}{3pt}
	\setlength{\belowdisplayskip}{3pt}
\begin{align}
\textbf{P3} : \hspace{-0.5mm}\max _{\mathbf{p}_\mathrm{s} ,t_{\mathrm{B},l^\mathrm{s}_{\bar{n}}}\hspace{-1mm},\atop t_{\mathrm{E},l^\mathrm{s}_{\bar{n}}}} & \sum\limits_{l^\mathrm{s}_{\bar{n}}\in \mathcal{L}_\mathrm{s}}\hspace{-2mm}\varphi_{\mathrm{B},l^\mathrm{s}_{\bar{n}}}\hspace{-1mm}\left( \mathbf{p}_\mathrm{s},t_{\mathrm{B},l^\mathrm{s}_{\bar{n}}}\hspace{-0.5mm}\right) \!-\!\sum\limits_{l^\mathrm{s}_{\bar{n}}\in \mathcal{L}_\mathrm{s}}\!\!\!\varphi_{\mathrm{E},l^\mathrm{s}_{\bar{n}}}\!\!\left( \mathbf{p}_\mathrm{s},t_{\mathrm{E},l^\mathrm{s}_{\bar{n}}}\right)  \\
\mathrm{s.t}.\quad \;\;\, & \sum\limits_{l^\mathrm{s}_{\bar{n}}\in \mathcal{L}_\mathrm{s}} p_{l^\mathrm{s}_{\bar{n}}} \leq \rho P_\mathrm{T}, \ p_{l^\mathrm{s}_{\bar{n}}} \geq p_\mathrm{th};\\
&t_{\mathrm{B},l^\mathrm{s}_{\bar{n}}},t_{\mathrm{E},l^\mathrm{s}_{\bar{n}}}>0, \forall l^\mathrm{s}_{\bar{n}} \in \mathcal{L}_\mathrm{s}.
\end{align}
\end{subequations}
  It is demonstrated that $ \textbf{P3} $ is convex for $ \mathbf{p}_\mathrm{s} $ or $\left(t_{\mathrm{B},l^\mathrm{s}_{\bar{n}}},t_{\mathrm{E},l^\mathrm{s}_{\bar{n}}}\right)  $ to be solved by the alternative optimization method.

According to \textbf{Lemma 1}, the optimal $ \left(t_{\mathrm{B},l^\mathrm{s}_{\bar{n}}}^{\mathrm{opt}},t_{\mathrm{E},l^\mathrm{s}_{\bar{n}}}^{\mathrm{opt}}\right) $ for given $ \mathbf{p}_\mathrm{s} $ can be derived as follows:
\begin{align}
	\setlength{\abovedisplayskip}{3pt}
	\setlength{\belowdisplayskip}{3pt}
\begin{split}
t_{\mathrm{B},l^\mathrm{s}_{\bar{n}}}^{\mathrm{opt}}=\left (1+\mathbf{a}_{-l^\mathrm{s}_{\bar{n}}}^\mathrm{T} \mathbf{p}_\mathrm{s}\right)^{-1}\hspace{-1mm};
t_{\mathrm{E},l^\mathrm{s}_{\bar{n}}}^{\mathrm{opt}}=\left (1+\mathbf{b}_{l^\mathrm{s}_{\bar{n}}}^\mathrm{T} \mathbf{p}_\mathrm{s}+c\right)^{-1}. \label{con:t_{B,E,l}}
\end{split}
\end{align}
Then, the optimal $ \mathbf{p}_\mathrm{s}^{\mathrm{opt}} $ can be obtained by solving problem $ \textbf{P4} $ given by 
\begin{subequations}
	\setlength{\abovedisplayskip}{3pt}
	\setlength{\belowdisplayskip}{3pt}
\begin{align}
\textbf{P4} : \; \max _{\mathbf{p}_\mathrm{s} }\hspace{-1mm} & \sum\limits_{l^\mathrm{s}_{\bar{n}}\in \mathcal{L}_\mathrm{s}}\hspace{-2mm}\varphi_{\mathrm{B},l^\mathrm{s}_{\bar{n}}}\left( \mathbf{p}_\mathrm{s},t_{\mathrm{B},l^\mathrm{s}_{\bar{n}}}^{\mathrm{opt}}\right)\hspace{-1mm} -\hspace{-2mm}\sum\limits_{l^\mathrm{s}_{\bar{n}}\in \mathcal{L}_\mathrm{s}}\hspace{-2mm}\varphi_{\mathrm{E},l^\mathrm{s}_{\bar{n}}}\left( \mathbf{p}_\mathrm{s},t_{\mathrm{E},l^\mathrm{s}_{\bar{n}}}^{\mathrm{opt}}\right)  \\
\mathrm{s.t}.\;  & \sum\limits_{l^\mathrm{s}_{\bar{n}}\in \mathcal{L}_\mathrm{s}} p_{l^\mathrm{s}_{\bar{n}}} \leq \rho P_\mathrm{T}, \  p_{l^\mathrm{s}_{\bar{n}}} \geq p_\mathrm{th}, \forall l^\mathrm{s}_{\bar{n}} \in \mathcal{L}_\mathrm{s}.
\end{align}
\end{subequations}
Because the objective function and constraint of problem $ \textbf{P4} $ are convex,  problem $ \textbf{P4} $ can be easily solved by convex problem solvers such as CVX.

Above all, problem $ \textbf{P2} $ can be solved by alternately updating $ \left(t_{\mathrm{B},l^\mathrm{s}_{\bar{n}}},t_{\mathrm{E},l^\mathrm{s}_{\bar{n}}}\right) $ and $ \mathbf{p}_\mathrm{s} $.

\setlength{\textfloatsep}{0.1cm}
\setlength{\floatsep}{0.1cm}
\begin{algorithm}[t]\small
	
	\caption{RMCG Based Alternative Optimization}
	\label{alg:1}
	\begin{algorithmic}[1]
		\STATE Initialize $ \mathbf{p}_{\mathrm{s},0} $ and $ \boldsymbol{\theta}_{0} $, 
		calculate $ \boldsymbol{\xi}_{0}=-\operatorname{grad}_{\boldsymbol{\theta}_{0}}f $ according to Eq. (\ref{con:grad}) and set $ i=0 $;
		\STATE \textbf{repeat}
		\STATE With given $ \mathbf{p}_{\mathrm{s},i} $ at the $i$-th iteration, calculate $ t_{\mathrm{B},l^\mathrm{s}_{\bar{n}},i+1} $ and $ t_{\mathrm{E},l^\mathrm{s}_{\bar{n}},i+1} $ with Eq. (\ref{con:t_{B,E,l}});
		\STATE Given $ t_{\mathrm{B},l^\mathrm{s}_{\bar{n}},i+1} $ and $ t_{\mathrm{E},l^\mathrm{s}_{\bar{n}},i+1} $, calculate $ \mathbf{p}_{\mathrm{s},i+1} $ by solving $ \textbf{P4} $;

		\STATE Given $ \boldsymbol{\theta}_{i} $, calculate Riemannian gradient $ \operatorname{grad}_{\boldsymbol{\theta}_{i}} f $ with Eq. (\ref{con:grad});
		
		\STATE Obtain Polak-Ribiere parameter $ \alpha_{i} $ with Eq. \eqref{con:Polak-Ribiere parameter} and then calculate the search direction $ \boldsymbol{\xi}_{i} $ with Eq. (\ref{con:search direction});
		\STATE  Find Armijo line search step size $ \beta_{i} $;
		\STATE Obtain $ \boldsymbol{\theta}_{i+1} $ according to Eq. (\ref{con:retraction});
		\STATE Update $ i=i+1 $;
		\STATE \textbf{until} convergence.
		\vspace{-2pt}
	\end{algorithmic} 
\end{algorithm}


\vspace{-0.25cm}
\subsection{Optimize $\boldsymbol{\theta}$ with given $\mathbf{p}_\mathrm{s}$}
\vspace{-0.1cm}
In the following, we calculate the optimal $ \boldsymbol{\theta}^{\mathrm{opt}} $ with given $\mathbf{p}_\mathrm{s}$. First, we denote $\omega _{l^\mathrm{s}_{\bar{n}},k} = \mathbf{f}_{l^\mathrm{s}_{\bar{n}}}^\mathrm{H} \mathbf{H}_{\mathrm{AB}}\mathbf{f}_{k} $,  $\boldsymbol{\mu}_{l^\mathrm{s}_{\bar{n}},k} = \operatorname{diag}\left(\mathbf{f}_{l^\mathrm{s}_{\bar{n}}}^\mathrm{H} \mathbf{H}_{\mathrm{RB}}\right) \mathbf{H}_{\mathrm{AR}} \mathbf{f}_{k}$, $ \zeta_{l^\mathrm{s}_{\bar{n}},\ddot{k}} =  \mathbf{h}_{\mathrm{AE},l^\mathrm{s}_{\bar{n}}} \mathbf{f}_{\ddot{k}} $, and $ \boldsymbol{\eta}_{l^\mathrm{s}_{\bar{n}},\ddot{k}} = \operatorname{diag}\left( \mathbf{h}_{\mathrm{RE},l^\mathrm{s}_{\bar{n}}}\right) \mathbf{H}_{\mathrm{AR}} \mathbf{f}_{\ddot{k}}  $. Thus, $ \tilde{R}_{\mathrm{B}} $ and $ R_{\mathrm{E}} $ can be re-written as $ \tilde{R}_{\mathrm{B}} = \sum\limits_{l^\mathrm{s}_{\bar{n}}\in \mathcal{L}_\mathrm{s}} \log _{2}\left(1+\bar{\gamma}_{\mathrm{B}, l^\mathrm{s}_{\bar{n}}}\right) $ and $ R_{\mathrm{E}} = \sum\limits_{l^\mathrm{s}_{\bar{n}}\in \mathcal{L}_\mathrm{s}} \log _{2}\left(1+\bar{\gamma}_{\mathrm{E}, l^\mathrm{s}_{\bar{n}}}\right) $, respectively, where
\begin{equation}
\setlength{\abovedisplayskip}{3pt}
\setlength{\belowdisplayskip}{3pt}	
\left\{\hspace{-2mm}
\begin{array}{lr}
\bar{\gamma}_{\mathrm{B}, l^\mathrm{s}_{\bar{n}}} \hspace{-1.5mm}=\hspace{-1mm} \frac{p_{l^\mathrm{s}_{\bar{n}}}\left|\omega _{l^\mathrm{s}_{\bar{n}}\!,l^\mathrm{s}_{\bar{n}}}+ \boldsymbol{\theta}^\mathrm{H} \boldsymbol{\mu}_{l^\mathrm{s}_{\bar{n}}\!,l^\mathrm{s}_{\bar{n}}} \right|^{2}}{\sum\limits_{ k \neq l^\mathrm{s}_{\bar{n}}\!, k \in \mathcal{L}_\mathrm{s}}\hspace{-4mm} p_{k}\left|\omega _{l^\mathrm{s}_{\bar{n}}\!,k}+ \boldsymbol{\theta}^\mathrm{H} \boldsymbol{\mu}_{l^\mathrm{s}_{\bar{n}}\!,k} \right|^{2}+\sigma_{\mathrm{B}}^{2}};\\
\bar{\gamma}_{\mathrm{E}, l^\mathrm{s}_{\bar{n}}} \hspace{-1.5mm}=\hspace{-1mm} \frac{p_{l^\mathrm{s}_{\bar{n}}}\left|  \zeta_{l^\mathrm{s}_{\bar{n}}\!,l^\mathrm{s}_{\bar{n}}}+ \boldsymbol{\theta}^\mathrm{H} \boldsymbol{\eta}_{l^\mathrm{s}_{\bar{n}}\!,l^\mathrm{s}_{\bar{n}}}  \right|^{2}}{\sum\limits_{ k \neq l^\mathrm{s}_{\bar{n}}\!, k \in \mathcal{L}_\mathrm{s}}\hspace{-4mm} p_{k}\hspace{-0.5mm}\left|  \zeta_{l^\mathrm{s}_{\bar{n}}\!,k}\hspace{-0.5mm}+ \boldsymbol{\theta}^\mathrm{H} \boldsymbol{\eta}_{l^\mathrm{s}_{\bar{n}}\!,k} \hspace{-0.5mm}\right|^{2}\hspace{-1mm}+\sigma_\mathrm{\ddot{z}}^2\hspace{-1mm}\sum\limits_{ l^\mathrm{z}_{\ddot{n}} \!\in\! \mathcal{L}_\mathrm{\ddot{z}}}\hspace{-1.5mm} \left|  \zeta_{l^\mathrm{s}_{\bar{n}}\!,l^\mathrm{z}_{\ddot{n}}}\hspace{-0.5mm}+ \boldsymbol{\theta}^\mathrm{H} \boldsymbol{\eta}_{l^\mathrm{s}_{\bar{n}}\!,l^\mathrm{z}_{\ddot{n}}}\! \right|^{2} \hspace{-1mm}+\sigma_{\mathrm{E}}^{2}}.
\end{array}
\right.
\end{equation}
Therefore, we can re-formulate problem $ \textbf{P1} $ as follows:
\begin{equation}
	\setlength{\abovedisplayskip}{3pt}
	\setlength{\belowdisplayskip}{3pt}
\textbf{P5}: \min _{\boldsymbol{\theta}}    f\hspace{-1mm}=\hspace{-0.5mm}-\hspace{-0.5mm}\left(\hspace{-0.5mm}\tilde{R}_{\mathrm{B}}\hspace{-0.5mm}-\hspace{-0.5mm}R_{\mathrm{E}}\hspace{-0.5mm}\right) \;\mathrm{s.t}. \left|\theta_{q}\right|  \hspace{-1mm}=\hspace{-1mm} 1, \forall q \hspace{-0.5mm}\in\hspace{-1mm} \left\lbrace\hspace{-0.5mm} 1,\dots, \hspace{-0.5mm}Q\hspace{-0.5mm}\right\rbrace.
\end{equation}
The unit modulus constraints $ \left|\theta_{q}\right|  = 1 $ form a complex circle manifold $ \mathcal{M} =\left\{\boldsymbol{\theta} \in \mathbb{C}^{Q \times 1}: \left|\theta_{1}\right| = \cdots =  \left|\theta_{Q}\right|= 1\right\} $. 
For given $ \boldsymbol{\theta}_{i} $ on the manifold $ \mathcal{M} $  at the $i$-th iteration, the tangent space, denoted by $ T_{\boldsymbol{\theta}_{i}} \mathcal{M} $, is given as follows:
\begin{equation}
	\setlength{\abovedisplayskip}{3pt}
	\setlength{\belowdisplayskip}{3pt}
T_{\boldsymbol{\theta}_{i}} \mathcal{M}=\left\{ \mathbf{v} \in \mathbb{C}^{Q \times 1} : \Re \left\{ \mathbf{v} \circ \boldsymbol{\theta}_{i}^{*} \right\}=\mathbf{0}_{Q}\right\},
\end{equation}
where $ \mathbf{v} $ is the tangent vector at $ \boldsymbol{\theta}_{i} $, $ \Re \left\{\cdot\right\}  $ is the real part of a complex number, $\circ$ represents
the Hadamard product between two matrices, and $ \left( \cdot\right) ^{*}  $ is the conjugate of a matrix. 
The Riemannian gradient $ \operatorname{grad}_{\boldsymbol{\theta}_{i}} f $ of function $ f $ at $ \boldsymbol{\theta}_{i} $ is the orthogonal projection of the Euclidean gradient $ \nabla_{\boldsymbol{\theta}_{i}} f $ onto
the tangent space $ T_{\boldsymbol{\theta}_{i}} \mathcal{M} $. Thus, we have
\begin{equation}
	\setlength{\abovedisplayskip}{3pt}
	\setlength{\belowdisplayskip}{3pt}
\operatorname{grad}_{\boldsymbol{\theta}_{i}} f=\nabla_{\boldsymbol{\theta}_{i}} f-\Re \left\{\nabla_{\boldsymbol{\theta}_{i}} f \circ \boldsymbol{\theta}_{i}^{*}  \right\} \circ \boldsymbol{\theta}_{i}, \label{con:grad}
\end{equation}
where $ \nabla_{\boldsymbol{\theta}_{i}} f $ is calculated by Eq. (\ref{con:Euclidean gradient}).

\setcounter{equation}{25}
Then, the search direction for RMCG method is updated by
\begin{equation}
	\setlength{\abovedisplayskip}{3pt}
	\setlength{\belowdisplayskip}{3pt}
	\begin{aligned}
\boldsymbol{\xi}_{i}=-\operatorname{grad}_{\boldsymbol{\theta}_{i}} f +\alpha_{i}\mathcal{T}_{\boldsymbol{\theta}_{i-1} \rightarrow \boldsymbol{\theta}_{i}}\left(\boldsymbol{\xi}_{i-1}\right),\label{con:search direction}
\end{aligned}
\end{equation}
where $ \boldsymbol{\xi}_{i-1} $ is the search direction at $ \boldsymbol{\theta}_{i-1} $, and $ \mathcal{T}_{\boldsymbol{\theta}_{i-1} \rightarrow \boldsymbol{\theta}_{i}}\left(\boldsymbol{\xi}_{i-1}\right) $ is a vector transport function mapping a tangent vector from one tangent space to another tangent space, and $ \alpha_{i} $ represents the Polak-Ribiere parameter \cite{zhong2022RMOCG}, which is given by
\begin{equation}
	\setlength{\abovedisplayskip}{3pt}
	\setlength{\belowdisplayskip}{3pt}
	\begin{aligned}
		\alpha_{i}=\left( \operatorname{grad}_{\boldsymbol{\theta}_{i}} f\right)^{\mathrm{H}} \frac{ \operatorname{grad}_{\boldsymbol{\theta}_{i}} f-\mathcal{T}_{\boldsymbol{\theta}_{i-1} \rightarrow \boldsymbol{\theta}_{i}}\left(\operatorname{grad}_{\boldsymbol{\theta}_{i-1}}f\right)  }{\Vert \operatorname{grad}_{\boldsymbol{\theta}_{i-1}} f\Vert ^{2}}. \label{con:Polak-Ribiere parameter}
	\end{aligned}
\end{equation}
 

According to the obtained search direction $ \boldsymbol{\xi}_{i} $ at $ \boldsymbol{\theta}_{i} $, we update $  \boldsymbol{\theta}_{i+1}  $ as follows:
\begin{equation}
	\setlength{\abovedisplayskip}{3pt}
	\setlength{\belowdisplayskip}{3pt}
\begin{aligned}
\boldsymbol{\theta}_{i+1}=\operatorname{unt}\left(\boldsymbol{\theta}_{i}+\beta_{i}\boldsymbol{\xi}_{i} \right),\label{con:retraction} 
\end{aligned}
\end{equation}
where $ \beta_{i} $ is the Armijo  line search step size\footnote{The step size is obtained by the Armijo line search strategy \cite{zhong2022RMOCG}. The key is to find the smallest integer $w \geq 0$ that satisfies $f\left(\boldsymbol{\theta}_{i} \right)- f\left(\boldsymbol{\theta}_{i}+\beta_{i-1}\nu^{w}\boldsymbol{\xi}_{i} \right) \geq \iota\beta_{i-1}\nu^{w}\Vert\operatorname{grad}_{\boldsymbol{\theta}_{i}} f\Vert^{2}$, where $\iota, \nu \in \left( 0,1\right) $ and $\beta_{i-1} > 0 $.  The step size at the $i$-th iteration is denoted by $\beta_{i}= \beta_{i-1}\nu^{w}$.} and $ \operatorname{unt}\left(\boldsymbol{\theta} \right) $  forms a vector whose elements are $\frac{\theta_{1}}{\vert \theta_{1}\vert},\ldots, \frac{\theta_{Q}}{\vert \theta_{Q}\vert} $. The optimal $ \boldsymbol{\theta}_{i} $ is obtained when the Riemannian gradient of the objective function is close to zero \cite{zhong2022RMOCG}. 

Based on the above-mentioned analysis, the entire algorithm for solving problem $ \textbf{P1} $ is summarized in Algorithm 1.
		
		

\vspace{-0.2cm}
\section{Numerical Results}\label{sec:Simulation Results}
\vspace{-0.1cm}
In this section, we present several numerical results to evaluate the performance of our proposed scheme. The parameters are set as follows: $ N=8 $, $ \mathbf{u}_{\mathrm{B}}=[0,0,20]^{\mathrm{T}} $ m, $ x_{\mathrm{R}}=2 $ m, $ y_{\mathrm{R}}=0 $ m, $ z_\mathrm{R}=20 $ m, $ P_\mathrm{T}=30 $ dBm, $\theta=0 $, $\varphi=-\pi/20 $, $\vartheta_{\mathrm{x}}=\vartheta_{\mathrm{y}}=\pi/4 $, $Q=150$, $ d_{\mathrm{y}}=d_{\mathrm{z}}=0.05 $ m, $ N_{\mathrm{A}}=4$, $ N_\mathrm{s}=N_\mathrm{\ddot{z}}=3$, $\rho=0.9$, and $\sigma_{\mathrm{B}}^{2}=  \sigma_{\mathrm{E}}^{2}=-20$ dBm. 
The low-order and high-order OAM-modes are  $\left\lbrace 0,+1,-1,-2 \right\rbrace $ and $\left\lbrace +2,-3,+3,+4 \right\rbrace $, respectively.
To show the effectiveness of our proposed scheme, we compare it with several other schemes: 1) RIS-OAM with EP: The phase shifts of RIS are required to be optimized with $\rho P_\mathrm{T}$ being equally allocated to each desired signal. 2) RIS-OAM without AN: All OAM-modes are used for signal transmission.
3) RIS-MIMO: AN is employed in RIS-MIMO secure communications. 4) Active RIS-OAM: Referring to \cite{zhi2022ActiveRIS}, the active RIS is deployed in OAM secure communications.

\begin{figure}[htbp]
	\centering
	\setlength{\abovecaptionskip}{-0.15cm}
	\centering
		\includegraphics[scale=0.44]{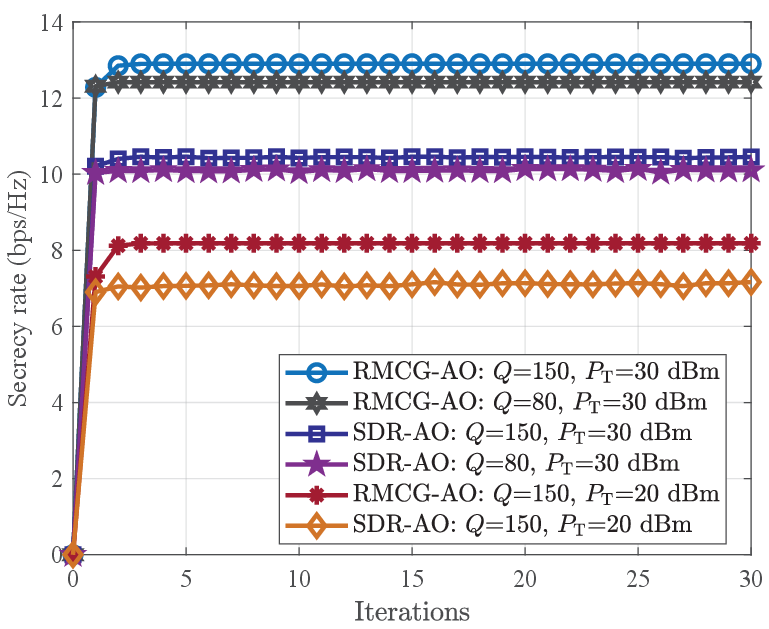}
		\caption{Secrecy rates versus the number of iterations.} \label{fig: iterations}
	\vspace{-5pt}
\end{figure}

Figure \ref{fig: iterations} shows the convergence of our developed RMCG-AO algorithm and the semidefinite relaxation (SDR) based AO algorithm with different  $P_\mathrm{T}$ and $Q$. It can be seen that the secrecy rates of these two algorithms increase rapidly to a fixed value as the iteration increases. Given $P_\mathrm{T}$ and $Q$, both RMCG-AO algorithm and SDR-AO algorithm reach convergence after about 3 iterations, but the former algorithm can achieve higher secrecy rates. The reason is that Gaussian randomization in the SDR-AO algorithm degrades the accuracy of the optimal solution.
Furthermore, the difference in the complexity of these two algorithms mainly lies in the complexity of calculating the phase shifts with RMCG method and SDR method. The complexity of RMCG method is $\mathcal{O}\left(N_{\mathrm{s} }^{2}Q^{2} \right)  $, while the complexity of SDR method is $\mathcal{O}\left(Q^{4.5} \right)   $ \cite{luo2010SDR}. Thus , the computational complexity of RMCG-AO algorithm is lower than that of SDR-AO algorithm when $ N_{\mathrm{s}} < Q^{\frac{5}{4}}$. Fig. 2 demonstrates the superiority of our proposed algorithm.

\begin{figure}[htbp]
	\centering
	\vspace{-13pt}
	\setlength{\abovecaptionskip}{-0.15cm}
	\centering		\includegraphics[scale=0.44]{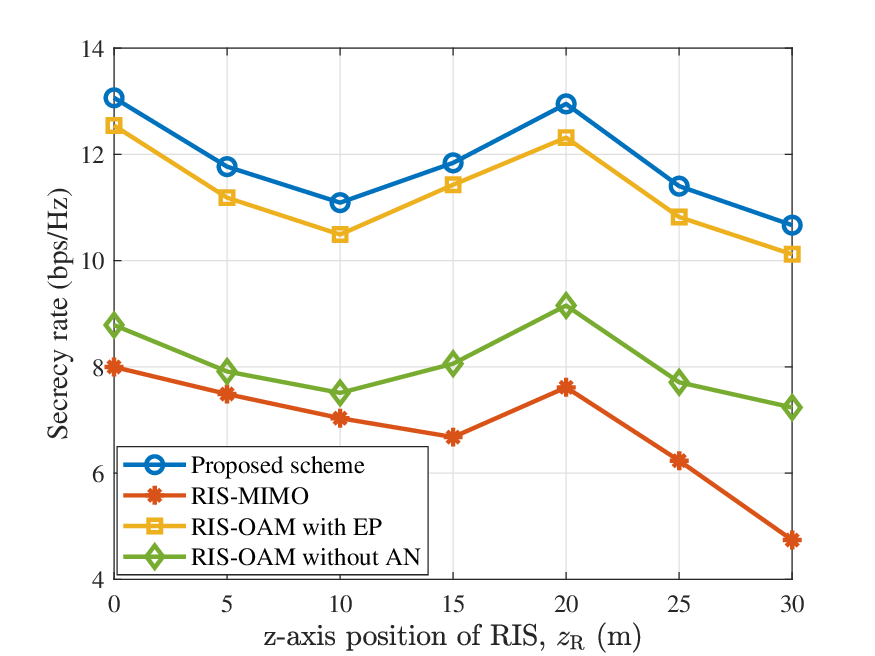}		
	\caption{Secrecy rates versus z-axis position of RIS,	    $z_\mathrm{R}$.} \label{fig: z_R}
	\vspace{-8pt}
\end{figure}
Figure \ref{fig: z_R} compares the secrecy rates of different schemes versus the z-axis position of RIS. The secrecy rates of our proposed scheme are higher than those of the other three schemes. Due to the low degrees of  freedom in LoS MIMO communications and the hollow structure of OAM beams, the secrecy rate of the RIS-MIMO scheme is lower than that of our proposed scheme. Due to a lack of the optimal transmit power allocation, the RIS-OAM with EP scheme has a lower secrecy rate than our proposed scheme. The RIS-OAM without AN scheme performs worse than our proposed scheme because Eve is not jammed by AN. Besides, the secrecy rates of all schemes first decrease and then increase as the RIS moves from Alice to Bob area ($ z_\mathrm{R} \leq 20$ m) along the z-axis. Also, the secrecy rate decreases obviously when the RIS is far away from Bob area ($ z_\mathrm{R} > 20$ m). The reason is that the channel strength is dependent on the product of the distance from Alice to the RIS and the distance from the RIS to Bob. This proves that a high secrecy rate can be achieved by deploying RIS near Alice or Bob. 

\begin{figure*}[htbp]
\centering 

	\setlength{\abovecaptionskip}{-0.15cm}
	\begin{minipage}[t]{0.3\linewidth}
	\centering
	\includegraphics[scale=0.38]{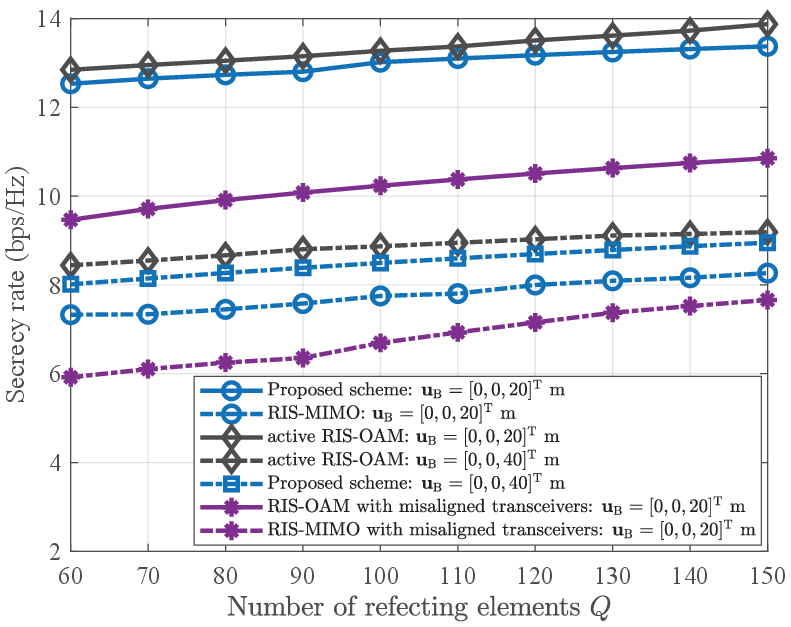}
	\caption{Secrecy rates versus the number\\
   of reflecting elements $Q$.} \label{fig:element}
\end{minipage}%
\begin{minipage}[t]{0.3\linewidth}

	\includegraphics[scale=0.38]{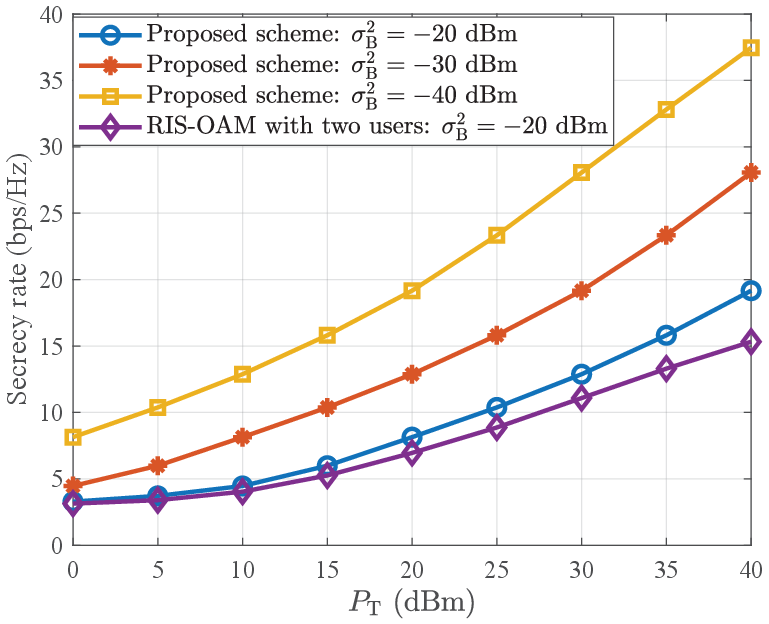}
	\caption{Secrecy rates versus the transmit power $P_{\mathrm{T}}$ with $\sigma_{\mathrm{B}}^{2}=  \sigma_{\mathrm{E}}^{2}$.} \label{fig:P_T-noise}
\end{minipage}
\begin{minipage}[t]{0.3\linewidth}

	\includegraphics[scale=0.38]{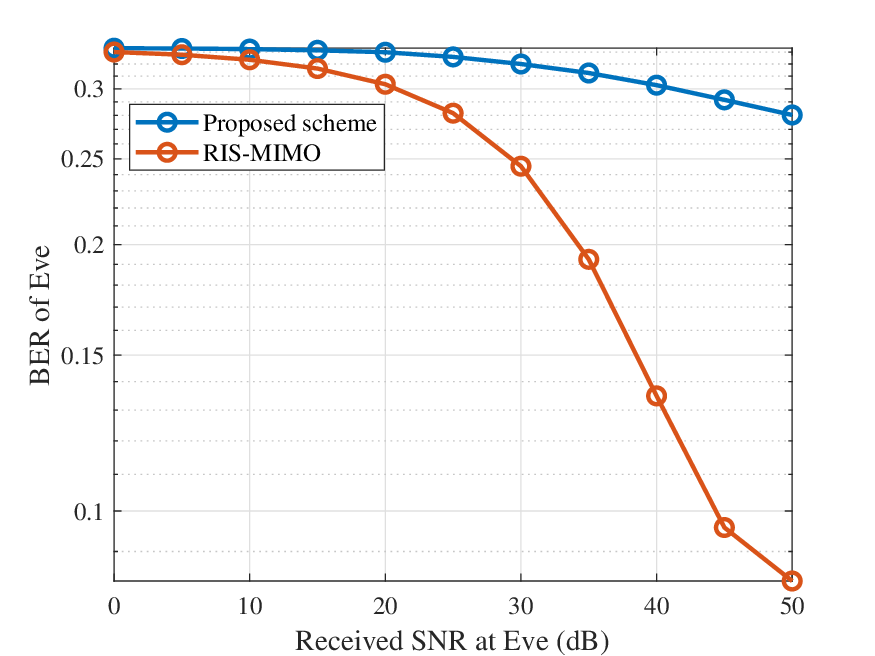}
	\caption{BERs of Eve versus the received SNR at Eve.} \label{fig:BER}
\end{minipage}
\vspace{-18pt}
\end{figure*}

Figure \ref{fig:element}  compares the secrecy rates of our proposed scheme, the RIS-MIMO scheme,  and the active RIS-OAM scheme versus the number of reflecting elements $Q$ with different $\mathbf{u}_\mathrm{B} $. It can be seen that the secrecy rates of all schemes increase as $Q$ increases. It verifies that designing the RIS’s phase shifts can enhance and weaken the desired signals received by Bob and Eve, respectively, thus increasing the secrecy rates. Because the orthogonality among OAM-modes is destroyed, the secrecy rates of our proposed scheme with aligned legitimate transceivers are higher than those of the RIS-OAM scheme with misaligned transceivers. With misaligned transceivers, the RIS-MIMO scheme still performs worse than the RIS-OAM scheme. The secrecy rates of the active RIS-OAM scheme are less than one bit higher than those of our proposed scheme with passive RIS. One reason is that  the signals suffer less double path fading effect in our proposed short-range secure communication scheme. Also, the direct LoS link between Alice and Bob for our proposed scheme plays a significant role in increasing the secrecy rate.

The impact of noise levels and the number of legitimate receivers on the secrecy rates of our proposed scheme is given in Fig. \ref{fig:P_T-noise}. As seen from Fig. \ref{fig:P_T-noise}, the secrecy rates of our proposed scheme increase with the decrease of noise power $\sigma_{\mathrm{B}}^{2}$. This is because that the SINR of Eve increases slower than that of Bob with the decrease of $\sigma_{\mathrm{B}}^{2}$, thus achieving a high secrecy rate for our proposed scheme in the low $\sigma_{\mathrm{B}}^{2}$ region. In addition, the secrecy rate of the RIS-OAM scheme with two users  is lower than that of our proposed scheme owing to the inter-user interference.


Figure \ref{fig:BER} illustrates the bit error rates (BERs) versus the received signal-to-noise ratio (SNR) at Eve of our proposed scheme and the RIS-MIMO scheme with quadrature phase shift keying (QPSK) modulation. As shown in Fig. \ref{fig:BER}, the BER of Eve for our proposed scheme is higher than that for the RIS-MIMO scheme. 
On the one hand, the initial transmit signals are susceptible to phase owing to the QPSK modulation and OAM at Alice. On the other hand, Eve mainly relies on the conventional MIMO to wiretap legitimate OAM signals. Hence, it is very difficult for Eve to successfully decompose initial transmit signals with conventional MIMO signal recovery methods in our proposed scheme, thus suffering from severe phase crosstalk. Therefore, Eve has high BERs for our proposed scheme. Figs. \ref{fig: z_R}, \ref{fig:element}, and \ref{fig:BER} show that our proposed scheme has better anti-eavesdropping results as compared with the conventional RIS-MIMO scheme.

\vspace{-0.1cm}
\section{Conclusions}\label{sec:Conclusion}
\vspace{-0.1cm}
In this paper, we proposed the RIS-assisted OAM scheme, where RIS and AN were leveraged to  jointly interfere with Eve, to enhance the secrecy rates in the presence of Eve for short-range secure communications. Introducing index modulation, several OAM-modes were randomly selected from low-order and high-order OAM-mode sets to carry desired signals and AN, respectively. The RMCG-AO algorithm was developed to jointly optimize the transmit power allocation and phase shifts of RIS to obtain the maximum secrecy rate. Numerical results have shown that our proposed scheme can significantly enhance the secrecy results as compared with the existing schemes in terms of secrecy rates and Eve's BERs. Future research will focus on active RIS-assisted OAM secure communication to steer the directions of OAM beams to mitigate the inter-mode crosstalk and reduce the divergence of OAM beams to extend the transmission distance, thus achieving highly reliable OAM  medium-long-distance secure communications.
\vspace{-0.1cm}
%
\bibliographystyle{IEEEtran}
\bibliography{References}

\begin{thebibliography}{10}
\providecommand{\url}[1]{#1}
\csname url@samestyle\endcsname
\providecommand{\newblock}{\relax}
\providecommand{\bibinfo}[2]{#2}
\providecommand{\BIBentrySTDinterwordspacing}{\spaceskip=0pt\relax}
\providecommand{\BIBentryALTinterwordstretchfactor}{4}
\providecommand{\BIBentryALTinterwordspacing}{\spaceskip=\fontdimen2\font plus
\BIBentryALTinterwordstretchfactor\fontdimen3\font minus
  \fontdimen4\font\relax}
\providecommand{\BIBforeignlanguage}[2]{{%
\expandafter\ifx\csname l@#1\endcsname\relax
\typeout{** WARNING: IEEEtran.bst: No hyphenation pattern has been}%
\typeout{** loaded for the language `#1'. Using the pattern for}%
\typeout{** the default language instead.}%
\else
\language=\csname l@#1\endcsname
\fi
#2}}
\providecommand{\BIBdecl}{\relax}
\BIBdecl

\bibitem{Beamforming_Power_Design_for_IRS}
F.~Shu, L.~Yang, X.~Jiang, W.~Cai, W.~Shi, M.~Huang, J.~Wang, and X.~You,
  ``Beamforming and transmit power design for intelligent reconfigurable
  surface-aided secure spatial modulation,'' \emph{IEEE Journal of Selected
  Topics in Signal Processing}, vol.~16, no.~5, pp. 933--949, Aug. 2022.

\bibitem{IRS_assisted_untrusted_NOMA}
D.~Wang, X.~Li, Y.~He, F.~Zhou, and Q.~Wu, ``Intelligent reflecting surface
  assisted untrusted {NOMA} transmissions: a secrecy perspective,''
  \emph{Science China Information Sciences}, vol.~66, 08 2023.

\bibitem{AN_Aided_Secure_MIMO_via_IRS}
S.~Hong, C.~Pan, H.~Ren, K.~Wang, and A.~Nallanathan, ``Artificial-noise-aided
  secure {MIMO} wireless communications via intelligent reflecting surface,''
  \emph{IEEE Transactions on Communications}, vol.~68, no.~12, pp. 7851--7866,
  Dec. 2020.

\bibitem{Index-Modulation}
L.~Liang, W.~Cheng, W.~Zhang, and H.~Zhang, ``Index-modulation embedded mode
  hopping for antijamming,'' \emph{IEEE Systems Journal}, vol.~16, no.~3, pp.
  3905--3916, Sep. 2022.

\bibitem{OAM_Secure_High_Speed_Communication}
I.~B. Djordjevic, ``Multidimensional {OAM}-based secure high-speed wireless
  communications,'' \emph{IEEE Access}, vol.~5, pp. 16\,416--16\,428, 2017.

\bibitem{Secure_Range_Dependent_Transmissio_With_OAM}
J.~Luo, S.~Wang, and F.~Wang, ``Secure range-dependent transmission with
  orbital angular momentum,'' \emph{IEEE Communications Letters}, vol.~23,
  no.~7, pp. 1178--1181, July 2019.

\bibitem{Physical_Layer_Secure_Communication_Using_OAM}
M.~A.~B. Abbasi, V.~Fusco, U.~Naeem, and O.~Malyuskin, ``Physical layer secure
  communication using orbital angular momentum transmitter and a single-antenna
  receiver,'' \emph{IEEE Transactions on Antennas and Propagation}, vol.~68,
  no.~7, pp. 5583--5591, July 2020.

\bibitem{ma2021OAMFDA}
J.~Ma, X.~Song, Y.~Yao, Z.~Zheng, X.~Gao, and S.~Huang, ``Secure transmission
  of radio orbital angular momentum beams based on the frequency diverse
  array,'' \emph{IEEE Access}, vol.~9, pp. 108\,924--108\,931, 2021.

\bibitem{zhang2022PLKOAM}
M.~Zhang, Z.~Ji, Y.~Zhang, P.~L. Yeoh, Z.~He, and Y.~Li, ``Physical layer key
  generation for secure {OAM} communication systems,'' \emph{IEEE Transactions
  on Vehicular Technology}, vol.~71, no.~11, pp. 12\,397--12\,401, Nov. 2022.

\bibitem{Transmit_Solutions_for_MIMO_AO}
Q.~Li, M.~Hong, H.-T. Wai, Y.-F. Liu, W.-K. Ma, and Z.-Q. Luo, ``Transmit
  solutions for {MIMO} wiretap channels using alternating optimization,''
  \emph{IEEE Journal on Selected Areas in Communications}, vol.~31, no.~9, pp.
  1714--1727, Sep. 2013.

\bibitem{zhong2022RMOCG}
K.~Zhong, J.~Hu, Y.~Cong, G.~Cui, and H.~Hu, ``{RMOCG}: A riemannian manifold
  optimization-based conjugate gradient method for phase-only beamforming
  synthesis,'' \emph{IEEE Antennas and Wireless Propagation Letters}, vol.~21,
  no.~8, pp. 1625--1629, Aug. 2022.

\bibitem{zhi2022ActiveRIS}
K.~Zhi, C.~Pan, H.~Ren, K.~K. Chai, and M.~Elkashlan, ``Active {RIS} versus
  passive {RIS}: Which is superior with the same power budget?'' \emph{IEEE
  Communications Letters}, vol.~26, no.~5, pp. 1150--1154, May 2022.

\bibitem{luo2010SDR}
Z.-q. Luo, W.-k. Ma, A.~M.-c. So, Y.~Ye, and S.~Zhang, ``Semidefinite
  relaxation of quadratic optimization problems,'' \emph{IEEE Signal Processing
  Magazine}, vol.~27, no.~3, pp. 20--34, May 2010.

\end{thebibliography}

\end{document}